\documentclass[journal]{IEEEtran}
\usepackage{amsmath, amssymb}
\usepackage{bm}
\usepackage{commath}
\usepackage{graphicx}
\usepackage[caption=false,font=footnotesize]{subfig}
\usepackage[hidelinks]{hyperref}
\usepackage{doi}
\usepackage[ruled,vlined,linesnumbered]{algorithm2e}
\SetAlCapNameFnt{\small}
\SetAlCapFnt{\small}

\input{variables.sty}

\title{Learning-Based Synthesis of Robust \\Linear Time-Invariant Controllers}

\author{Marc-Antoine Beaudoin,
	Benoit Boulet,~\IEEEmembership{Senior Member, IEEE}%
	\thanks{The authors are with the Intelligent Automation Lab of the Centre for Intelligent Machines and the Department of Electrical and Computer Engineering, McGill University, Montreal, QC, Canada, H3A 0E9 e-mail: (ma.beaudoin@mail.mcgill.ca, benoit.boulet@mcgill.ca).}%
	\thanks{Manuscript received February 25, 2022; revised April 2, 2022; accepted April 19, 2022. This work was supported by Mitacs and Quebec's Fonds de recherche Nature et technologies.}}%

\begin{document}
\maketitle

\begin{abstract}
	Recent advances in learning for control allow to synthesize vehicle controllers from learned system dynamics and maintain robust stability guarantees. However, no approach is well-suited for training robustly-stabilizing linear time-invariant (LTI) controllers using arbitrary learned models of the dynamics. This article introduces a method to do so. It uses a robust control framework to derive robust stability criteria. It also uses simulated policy rollouts to obtain gradients on the controller parameters, which serve to improve the closed-loop performance. By formulating the stability criteria as penalties with computable gradients, they can be used to guide the controller parameters toward robust stability during gradient descent. The approach is flexible as it does not restrict the type of learned model for the simulated rollouts. The robust control framework ensures that the controller is already robustly stabilizing when first implemented on the actual system and no data is yet collected. It also ensures that the system stays stable in the event of a shift in dynamics, given the system behavior remains within assumed uncertainty bounds. We demonstrate the approach by synthesizing a controller for simulated autonomous lane-change maneuvers. This work thus presents a flexible approach to learning robustly stabilizing LTI controllers that takes advantage of modern machine learning techniques.
\end{abstract}


\section{Introduction}
\IEEEPARstart{M}{achine} learning can be used to improve the performance of controllers for intelligent vehicles. Typically the performance increase is enabled by a reduction in the modeling uncertainty. In fact, the fundamental tradeoff between robustness and performance in control engineering~\cite{boulet_fundamental_2007} reveals how modeling uncertainty can oftentimes be the limiting factor in system performance. However, it can be hard to maintain theoretical guarantees of stability and safety when incorporating learning components in control methods.\\

For vehicular control, neural network controllers can be trained with deep reinforcement learning and reach interesting levels of robustness~\cite{amini_learning_2020,lin_comparison_2021}. Domain randomization~\cite{tobin_domain_2017,andrychowicz_learning_2020} is often used to increase robustness. The idea is to randomly alter the physics of the simulated environment used by the learning algorithm, which better prepares the controller to be deployed in the real world. But it remains that whenever possible, it is preferable to maintain theoretical guarantees, and domain randomization does not. \\

In recent years, several lines of research presented successful approaches to theoretical guarantees. A safety framework~\cite{fisac_general_2019} based on Hamilton-Jacobi reachability analysis~\cite{chen_hamiltonjacobi_2018} can be used to overwrite any learning controller in the event that the system is suspected to exit a given safe set. As more data is gathered, the safety framework gradually allows the learning controller to come closer to the unsafe region, as it becomes more certain of the system dynamics through learning with Gaussian processes (GP)~\cite{rasmussen_gaussian_2006,liu_gaussian_2018}. Another approach~\cite{berkenkamp_safe_2017} consists of using Lyapunov theory to characterize the system's region of attraction (ROA), and tune a neural network (NN) controller with gradient descent while considering the constraints imposed by the ROA. Here again, GPs are used to learn the system dynamics. The size of the ROA and the controller performance both gradually increase as a result of controller tuning and data collection. In two closely related approaches~\cite{taylor_episodic_2019,taylor_learning_2020}, a quadratic programming controller is used to adjust a nominal controller's output as to enforce a safety constraint. In~\cite{taylor_episodic_2019} the safety constraint is defined using a control Lyapunov function, and in~\cite{taylor_learning_2020}, a control barrier function (CBF). In these approaches the safety is contingent on having an accurate model of the system dynamics, so a NN is trained from collected data and used to improve the model's accuracy. A time-varying CBF is introduced in~\cite{huang_stability_2021} to enforce stability for vehicular control. In~\cite{beaudoin_structured_2022} learned control barrier functions are used to keep vehicles within state constraints despite modeling uncertainty, and the safe uncertainty-learning principle is introduced to guide the process of uncertainty estimation. In~\cite{helwa_provably_2019} researchers devised an outer-loop controller to adjust the output of a nominal controller. The model uncertainty estimates are updated in real time using the variance of a GP regression in order to better adjust the controller's aggressiveness with respect to the local system configuration. Researchers in~\cite{dean_sample_2019} use machine learning to estimate a linear system model along with error bounds, from which they can obtain a robust linear controller. The method can also be used to identify systems having linear dynamics with respect to nonlinear features of states and inputs~\cite{mania_active_2022}. It can also be used to design robust linear controllers for systems with complex high-dimensional sensorial inputs like camera-based autonomous vehicles~\cite{dean_robust_2020}, thereby taking steps toward formal guarantees even for the most complex problems. Finally, model predictive control (MPC) methods---which are often used for vehicular control~\cite{paden_survey_2016,brudigam_stochastic_2022}---can also benefit from learning~\cite{hewing_learning-based_2020}. In~\cite{hewing_cautious_2020} researchers learn the system dynamics with GPs, which provides an uncertainty description than can be used in a stochastic MPC formulation in order to obtain probabilistic guarantees of keeping the system within state constraints.\\

In this study, we introduce a control learning method that is appropriate for robustly-stabilizing linear time-invariant (LTI) controllers, which typically have the state-space representation 
\begin{align}
	\dbxc &= \Ak \bxc + \Bk \be, \\
	\bu &= \Ck \bxc + \Dk \be,
\end{align}
where $\be$ is the error signal fed to the controller, $\bxc$ are the controller states, and $\bu$ are the controller output signals. A time-proven method to synthesize such a controller is $\Hinf$ loop shaping within a robust control framework~\cite{zhou_essentials_1998}---see~\cite{hyde_application_1993,boulet_uncertainty_1997} for instances. To do so, one assumes a nominal linear system model along with uncertainty bounds, defines performance criteria in terms of filters on the input and output signals, and solves a norm-minimization problem. However, $\Hinf$ loop shaping can be difficult to implement in practice. Notably, the method is restricted to linear system models, so it can be hard to optimize closed-loop performance when the system is best described with a nonlinear model. As a result, the method is ill-suited for taking advantage of system identification through modern machine learning techniques, as the learned models are often nonlinear. Finally, it can be hard to translate the desire to track specific temporal reference trajectories into designs filters. Thus the objective in this research is to develop a LTI controller synthesis method that can take advantage of nonlinear learned models while ensuring robust closed-loop stability through a robust control framework.\\

Such a method does not exists in the literature. The approaches in~\cite{fisac_general_2019,berkenkamp_safe_2017,taylor_episodic_2019,taylor_learning_2020,helwa_provably_2019} all consist of adjusting a base controller output according to a safety criteria computed in real time, but they do not tune the base controller parameters. Moreover, the approaches of~\cite{fisac_general_2019,berkenkamp_safe_2017} suffer from the curse of dimensionality, which makes them ill-suited for multi-input multi-output systems with several states---a typical application for LTI controllers. The approach of~\cite{dean_sample_2019} could be used to synthesize such controllers, but it assumes a linear system model, which defeats the objective of this work. Researchers in~\cite{fiedler_learning-enhanced_2021} use a robust control framework to ensure robust stability. By learning a probabilistic dynamical model with GPs, they reduce the modeling uncertainty by translating the GP posterior variance into the robust control framework, and ultimately increase the closed-loop performance. However, this approach is still a norm-minimization problem with design filters. Another approach~\cite{kretchmar_robust_2001,jin_stability-certified_2020} is to attach a NN to a LTI controller. But in this work we aim to only tune the LTI controller without requiring a NN.\\

This paper introduces the method outlined in Algorithm~\ref{alg:RobStab}. To obtain formal guarantees of robust stability, we define a nominal linear system model $G(s)$ and uncertainty bounds. Then robust stability criteria are derived from a robust control framework. Also, an initial controller $\Ki$ is obtained from $\Hinf$ loop shaping. Then, the controller is tuned with gradients obtained from simulated rollouts using any nominal model $\fn(\bx,\bu)$, thereby obtaining the tuned controller $\Kt$. The dynamics $\fn(\bx,\bu)$ can be the same as $G(s)$, but can also be a nonlinear model. During the training, the robust stability criteria are enforced through the gradients. Finally, the tuned controller $\Kt$ can be adjusted to the actual system dynamics, thereby obtaining the final adjusted controller $\Ka$. This is done by implementing $\Kt$ on the actual system, collecting data, learning a system model $\fl(\bx,\bu)$, and then retraining the controller from the same gradient descent method. The same robust stability criteria are enforced during this second controller tuning. \\

Tuning controllers from simulated rollouts was done in \textsc{pilco}~\cite{deisenroth_pilco:_2011}---a successful control learning approach that inspired part of the proposed method, notably the performance gradients presented in Section~\ref{sec:perf_grad}. \textsc{Pilco} is a powerful method to tune controllers for physical systems since it requires comparatively few data points to improve closed-loop performance, as we demonstrated in~\cite{beaudoin_improving_2022} when tuning a gearshift controller for a multi-speed transmission. However, \textsc{pilco} does not explicitly enforce robust stability though constraints in the optimization problem. Instead, \textsc{pilco} propagates state uncertainty during rollouts, which is used to penalize controller behaviors leading to regions of the state space with high uncertainty. Because of the robust stability criteria, our proposed method can rely on deterministic policy rollouts, which is interesting given that uncertainty propagation is not a trivial problem.\\

The proposed method synthesizes a controller that is already robustly stabilizing when it is first implemented on the actual system. This is not the case for some of the methods outlined above, such as ~\cite{taylor_episodic_2019,taylor_learning_2020}. Moreover, the controller remains stable even in the event of a sudden change in system dynamics, given that this change remains within assumed uncertainty bounds. By design, the learning methods reviewed above all rely on the assumption that the system dynamics are fixed. This allows to learn the uncertainty bounds from data, and increase the closed-loop performance. But it leaves no formal guarantee for the system safety when a sudden change in dynamics occur, which could be a concern for certain application. On the contrary, if for a given application it is judged that the fixed uncertainty bounds are too conservative, our methods still makes it possible to learn the bounds such as in~\cite{fiedler_learning-enhanced_2021}, and retrain the controller. \\

In summary, the learning-based controller synthesis method introduced in this article maintains theoretical guarantees of robust stability during learning, which other methods such as domain randomization and \textsc{pilco} do not. Because the method does not suffer from the curse of dimensionality, it is well-suited for multi-input multi-output systems and for linear controllers with a large number of internal states. The key constituents of this method are the use of a robust control framework to derive robust stability criteria, and gradient-based parameter optimization from simulated deterministic policy rollouts. The robust stability criteria are employed during the gradient-based optimization to maintain robust stability. Because the method is based on simulated policy rollouts, it is well-suited for trajectory-tracking problems where the performance criteria are hard to express as design filters. Moreover, it allows to use any learned model of the system dynamics such as neural networks and Gaussian processes. Finally, the robust control framework guarantees robust stability for a set of dynamical systems defined with uncertainty bounds. As a result, the controller is robustly stabilizing even before data can be collected, and it remains robustly stabilizing even if the system dynamics vary within the chosen uncertainty bounds. \\

Section~\ref{sec:robust} presents the robust control framework used in this work. Section~\ref{sec:prop_method} introduces the proposed controller learning method. An example problem is presented in Section~\ref{sec:example}, namely synthesizing a controller for a lane-change maneuver with an autonomous vehicle. The results are discussed in Section~\ref{sec:results}.

\section{Robust control framework} \label{sec:robust}
This section reviews concepts of robust control---a more detailed treatment is available in~\cite{zhou_essentials_1998}---and it introduces the robust control framework used in Algorithm~\ref{alg:RobStab}. In this article, the notation $\Vert G(s) \Vert_\infty$ refers to the $\infty$-norm of a linear system $G(s)$ and is defined as $\Vert G(s) \Vert_\infty := \sup_\omega \bar{\sigma}\big(G(j\omega)\big)$, where $\bar{\sigma}(M)$ denotes the maximum singular value of a transfer matrix $M$. The space $\mathcal{R}\Hinf$ is the space of all proper and real rational stable transfer functions. \\

\begin{figure}
	\centering
	\includegraphics[width=0.7\linewidth,trim={0.2cm 22.6cm 12.6cm 0.2cm},clip]{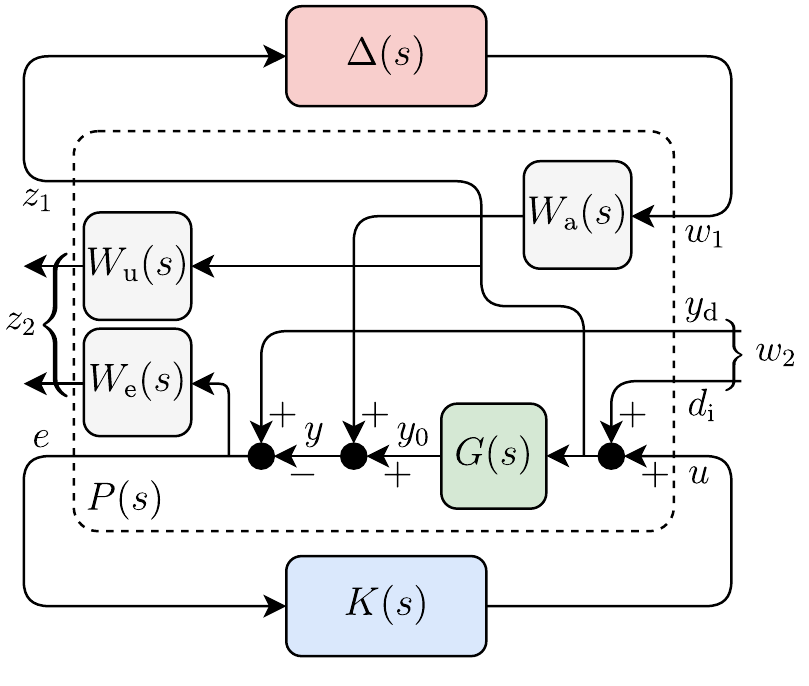}
	\caption{Robust control framework used to derive the robust stability criteria.}
	\label{fig:control}
\end{figure}

A typical robust control problem is formulated based on system models such as that of Figure~\ref{fig:control}. The goal is to find a stabilizing controller $K(s)$ that maximizes the system performance, which means minimizing $\Vert \Tztwt(s) \Vert_\infty$, where $\Tztwt(s)$ is the transfer function between the inputs $w_2$ and the outputs $z_2$. Design filters are used to weight the penalty on the various signals of $z_2$ over frequency ranges of interest. In Figure~\ref{fig:control}, $\Wu(s)$ penalizes the control signal and $\We(s)$, the error signal. Because the actual system dynamics is not perfectly known, it is assumed to be within a set of perturbed plants $\mathcal{G}$. Thus, nominal stability is not sufficient; $K(s)$ must stabilize the system for all possible plants in $\mathcal{G}$. In Figure~\ref{fig:control}, additive uncertainty is used to describe $\mathcal{G} := \{ G(s) + \Wa(s)\Delta(s) \}$, where $\Delta(s) \in \mathcal{R}\Hinf$ with $\Vert \Delta(s) \Vert_\infty < 1$, $G(s)$ is the nominal plant dynamics, and $\Wa(s)$ is a design filter bounding the uncertainty. Also commonly used are output multiplicative uncertainty, where $\mathcal{G} := \{(I+\Wm(s)\Delta(s))G(s)\}$, and input multiplicative uncertainty, where $\mathcal{G} := \{G(s)(I+\Wi(s)\Delta(s))\}$. From the small gain theorem~\cite{zhou_essentials_1998}, if $\Delta(s) \in \mathcal{R}\Hinf$ and $\Tzowo(s) \in \mathcal{R}\Hinf$, the system of Figure~\ref{fig:control} is robustly stabilizing if and only if ${\Vert \Delta(j\omega) \Tzowo(j\omega) \Vert < 1 \ \forall \omega}$. And because $\Vert \Delta(s) \Vert_\infty < 1$ by design, robust stability is guaranteed if $K(s)$ stabilizes $G(s)$ and makes $\Vert \Tzowo(s) \Vert_\infty \leq 1$.

\begin{figure}
	\centering
	\includegraphics[width=0.52\linewidth,trim={0.2cm 25.9cm 15.0cm 0cm},clip]{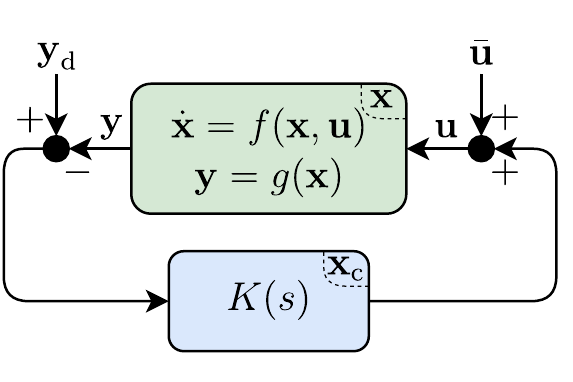}
	\caption{Closed-loop system used during policy rollouts.}
	\label{fig:control_nonlinear}
\end{figure}

\section{Controller learning method} \label{sec:prop_method}
The proposed approach is outlined in Algorithm~\ref{alg:RobStab}. It requires to define a nominal system model $G(s)$ along with uncertainty bounds such that all possible system dynamics are contained in a set $\mathcal{G}$. Using the robust control framework of Section~\ref{sec:robust}, this set is used to define robust stability criteria. The method also requires to define possibly nonlinear system models with the form
\begin{align}
	\dot{\bx} &= f(\bx,\bu), \\
	\by &= g(\bx),
\end{align} 
with states $\bx$, inputs $\bu$, and outputs $\by$. These system models are to be used during the simulated policy rollouts under the closed-loop configuration of Figure~\ref{fig:control_nonlinear}. The simulations consist of tracking $N_\mathrm{t}$ trajectories in rollouts of $T$ time steps. The while loops starting at lines~\ref{algl:while}~and~\ref{algl:while2} tune the controllers by solving the constrained optimization problem 
\begin{align}
	\argmin_{\bphi} \,  \cp(\bphi) &= \argmin_{\bphi} \sum_{n_\mathrm{t}=1}^{N_\mathrm{t}} \sum_{t=0}^{T} c(\by_{[t]}), \\
	\mathrm{s.t.} \quad & K(s) \ \mathrm{stabilizes} \ G(s), \label{eq:const_stab}\\
	& \Vert \Tzowo(s) \Vert_\infty \leq 1, \label{eq:const_rob}
\end{align}
where $\bphi$ contains the control parameters to be tuned, namely the entries of the matrices $\Ak$, $\Bk$, $\Ck$, and $\Dk$. The two constraints of Equations~\eqref{eq:const_stab}~and~\eqref{eq:const_rob} are turned into penalties with easily computable gradients. As such, a more general cost function is 
\begin{equation}
	J(\bphi) = \cp(\bphi) + \xis \cs(\bphi) + \xir \crob(\bphi), \label{eq:cost}
\end{equation}
where $\cp$ is the performance cost, $\cs$ is the nominal stability penalty, and $\crob$ is the robust stability penalty. The controller parameters $\bphi$ are iterated through either a simple stochastic gradient descent scheme---where $\bphi_{[i+1]} \leftarrow \bphi_{[i]} + \alpha \nabla_{\bphi} J(\bphi)$ and $\alpha$ is the learning rate---or any other suitable optimization algorithm. Constants $\xis,\xir \in \mathbb{R}^{+}$ are used to weight the penalties and help the gradient descent method. The next sections demonstrate how to obtain the gradients of the performance cost and penalties.

\begin{algorithm}
	\caption{Learning-based synthesis of robust controllers} \label{alg:RobStab}
	\small
	\KwResult{Learned controller parameters $\bphi = \{\Ak, \Bk, \Ck, \Dk \}$.}
	Define nominal linear system dynamics $G(s)$.\\
	Define set of all possible system dynamics $\mathcal{G}$, bounding the uncertainty on $G(s)$.\\
	Define weighting functions $\We(s)$ and $\Wu(s)$; obtain $\Ki(s)$ from $\Hinf$ loop shaping. \\
	Define nominal (possibly nonlinear) system dynamics $\fn(\bx,\bu)$. \\
	Initialize $\Kt(s) = \Ki(s)$.\\
	\While{$\Kt(s)$ not tuned}{ \label{algl:while}
		Simulate a policy rollout with $\Kt(s)$ on $\fn(\bx,\bu)$ as per Figure~\ref{fig:control_nonlinear}.\\
		Compute the total gradients $\nabla_{\bphi} J(\bphi) = \nabla_{\bphi}\cp(\bphi) + \xis \nabla_{\bphi}\cs(\bphi) + \xir \nabla_{\bphi}\crob(\bphi)$. \\
		Iterate the parameters $\bphi$ with $\nabla_{\bphi} J(\bphi)$.
	} 
	Initialize $\Ka(s) = \Kt(s)$. \\
	\While{$\Ka(s)$ not learned}{
		Implement $\Ka(s)$ on actual system, collect dataset $\mathcal{D}$. \label{algl:imp}\\
		Learn a model $\fl(\bx,\bu)$ for the system dynamics from $\mathcal{D}$.\\
		\While{$\Ka(s)$ not tuned}{ \label{algl:while2}
			Simulate a policy rollout with $\Ka(s)$ on $\fl(\bx,\bu)$ as per Figure~\ref{fig:control_nonlinear}.\\
			Compute the total gradients $\nabla_{\bphi} J(\bphi) = \nabla_{\bphi}\cp(\bphi) + \xis \nabla_{\bphi}\cs(\bphi) + \xir \nabla_{\bphi}\crob(\bphi)$. \\
			Iterate the parameters $\bphi$ with $\nabla_{\bphi} J(\bphi)$.
		}
	}
\end{algorithm}

\subsection{Performance gradients} \label{sec:perf_grad}
In discrete time, the system of Figure~\ref{fig:control_nonlinear} transitions with
\begin{align}
	\by_{[t]} &= g(\bx_{[t]}), \\
	\bx_{[t]} &= \bx_{[t-1]} + \Ts f(\bx_{[t-1]},\bu_{[t-1]}), \\
	\bu_{[t-1]} &= \Ck {\bxc}_{[t-1]} + \Dk ({\byd}_{[t-1]} - \by_{[t-1]}) + {\bbu}_{[t-1]}, \\
	{\bxc}_{[t-1]} &= (I + \Ts \Ak) {\bxc}_{[t-2]}+ \Ts \Bk ( {\byd}_{[t-2]} - \by_{[t-2]} ),
\end{align}
where $\Ts$ is the sampling period, $\bbu$ is a nominal (feedforward) control command, and $\byd$ is the desired output trajectory. The performance gradients are obtained by applying the chain rule on the results of the simulated policy rollout
\begin{align}
	\dod{\cp(\bphi)}{\bphi}  &= \sum_{n_\mathrm{t}=1}^{N_\mathrm{t}} \sum_{t=0}^{T} \dod{c(\by_{[t]})}{\bphi}, \\
	\dod{c(\by_{[t]})}{\bphi} &= \dod{c(\by_{[t]})}{\by_{[t]}} \dod{\by_{[t]}}{\bphi}. \label{eq:perf_grad}
\end{align}
The first term on the right-hand side of Equation~\eqref{eq:perf_grad} depends on the chosen cost function, the other term can be further expanded as
\begin{align}
	&\dod{\by_{[t]}}{\bphi} = \dod{g(\bx_{[t]})}{\bx_{[t]}} \dod{\bx_{[t]}}{\bphi}, \\
	&\dod{\bx_{[t]}}{\bphi} = \dod{\bx_{[t-1]}}{\bphi} + \Ts \dod{}{\bphi}f(\bx_{[t-1]},\bu_{[t-1]}), \\
	&\dod{}{\bphi}f(\bx_{[t-1]},\bu_{[t-1]}) = \nonumber \\
	&\qquad \dpd{f}{\bx_{[t-1]}} \dod{\bx_{[t-1]}}{\bphi} + \dpd{f}{\bu_{[t-1]}} \dod{\bu_{[t-1]}}{\bphi}, \\
	&\dod{\bu_{[t-1]}}{\bphi} = \Ck \dod{{\bxc}_{[t-1]}}{\bphi} + (I)_{ij}({\bxc}_{[t-1]})_k\dod{\Ck}{\bphi} \nonumber \\
	&\qquad -\Dk \dod{\by_{[t-1]}}{\bphi} + (I)_{ij}({\byd}_{[t-1]}-{\by}_{[t-1]})_k\dod{\Dk}{\bphi}, \label{eq:du}\\
	&\dod{{\bxc}_{[t-1]}}{\bphi} = (I+\Ts \Ak)\dod{{\bxc}_{[t-2]}}{\bphi} \nonumber \\
	&\qquad + \Ts(I)_{ij}({\bxc}_{[t-2]})_k\dod{\Ak}{\bphi} - \Ts \Bk \dod{{\bx}_{[t-2]}}{\bphi} \nonumber \\
	&\qquad  + \Ts(I)_{ij}({\byd}_{[t-2]} - {\by}_{[t-2]})_k\dod{\Bk}{\bphi}, \label{eq:dxc}
\end{align}
where $\od{\bx_{[t-1]}}{\bphi}$ and $\od{{\bxc}_{[t-2]}}{\bphi}$ are obtained from the previous time steps. In Equations~\eqref{eq:du} and \eqref{eq:dxc}, the Einstein summation convention is used for higher-order tensor manipulations. For instance, the product of a matrix $A_{ij}$ with a vector $\bx_k$ would be a third-order tensor $C_{ijk}$.

\subsection{Nominal stability gradients}
In the configuration of Figure~\ref{fig:control}, $K(s)$ stabilizes $G(s)$ if and only if the following system is stable
\begin{align}
	\begin{bmatrix}
		y(s) \\
		u(s)
	\end{bmatrix} &=
	\begin{bmatrix}
		S_1(s) & S_2(s) \\
		S_3(s) & S_4(s)
	\end{bmatrix}
	\begin{bmatrix}
		\yd(s) \\
		\di(s)
	\end{bmatrix}, \label{eq:T_stab} \\
	S_1(s) &= G(s)K(s)\big( I + G(s)K(s) \big)^{-1}, \label{eq:S1} \\
	S_2(s) &= G(s)\big( I + K(s)G(s) \big)^{-1}, \\
	S_3(s) &= K(s)\big( I + G(s)K(s) \big)^{-1}, \\
	S_4(s) &= K(s)G(s)\big( I + K(s)G(s) \big)^{-1}. \label{eq:S4}
\end{align}
For transfer functions with the state space realizations
\begin{equation}
	G(s)=
	\left[
	\begin{array}{c|c}
		\Ag & \Bg \\
		\hline
		\Cg & 0
	\end{array}
	\right], \quad 
	K(s)=
	\left[
	\begin{array}{c|c}
		\Ak & \Bk \\
		\hline
		\Ck & \Dk
	\end{array}
	\right]
\end{equation}
it suffices to verify that the eigenvalues $\lambda$ of the four matrices
\begin{align}
	M_1 &= \begin{bmatrix}
		\Ag & \Bg \Ck & \Bg \Dk \Cg\\
		0 & \Ak & -\Bk \Cg \\
		0 & \Bg \Ck & \Ag - \Bg \Dk \Cg
	\end{bmatrix}, \\
	M_2 &= \begin{bmatrix}
		\Ag - \Bg \Dk \Cg & -\Bg \Ck \\
		\Bk \Cg & \Ak
	\end{bmatrix}, \\
	M_3 &= \begin{bmatrix}
		\Ak & -\Bk \Cg \\
		\Bg \Ck & \Ag - \Bg \Dk \Cg
	\end{bmatrix}, \\
	M_4 &= \begin{bmatrix}
		\Ak & \Bk \Cg & 0 \\
		0 & \Ag - \Bg \Dk \Cg & -\Bg \Ck \\
		0 & \Bk \Cg & \Ak
	\end{bmatrix}, 
\end{align}
are all in the open left-half plane. In effect, these are the $A$ matrices of the state space realizations of the transfer functions in Equations~(\ref{eq:S1})--(\ref{eq:S4}). In the learning algorithm, a penalty is applied whenever the system is unstable. The penalty for the nominal stability is therefore defined as
\begin{equation}
	\cs(\bphi) = \sum_{i=1}^{4} \max \Big\{ 0, \max_j \big\{\Re\big( \lambda_j (M_i) \big) \big\} \Big\},
\end{equation}
where $\Re(\cdot)$ designates the real part of a complex number. The gradient of $\cs$ is easily obtainable with automatic differentiation~\cite{baydin_automatic_2018}. In this work, we used TensorFlow~\cite{abadi_tensorflow_2016}. 

\subsection{Robust stability gradients} \label{sec:rob_grad}
A state space representation for $\Tzowo(s)$ is obtained with the linear fractional transformation ${\Tzowo(s) = \mathcal{F}_L [\Pred(s),K(s)]}$, where $\Pred(s)$ represents ${[z_1,e]^\top = \Pred(s)[w_1,u]^\top}$, and is partitioned as follows
\begin{equation}
	\Pred(s) = \left[ \begin{array}{c|c c}
		A & B_1 & B_2 \\
		\hline
		C_1 & D_{11} & D_{12} \\
		C_2 & D_{21} & 0
	\end{array} \right].
\end{equation}
Then,
\begin{align}
	&\Tzowo(s) = 
	\left[
	\begin{array}{c|c}
		\bar{A} & \bar{B} \\
		\hline
		\bar{C} \rule{0pt}{2.5ex} & \bar{D}
	\end{array} 
	\right] = \nonumber \\
	&
	\left[
	\begin{array}{c c|c}
		A + B_2 \Dk C_2 & B_2 \Ck & B_1 + B_2 \Dk D_{21} \\
		\Bk C_2 & \Ak & \Bk D_{21} \\
		\hline
		C_1 + D_{12} \Dk C_2 & D_{12} \Ck & D_{11} + D_{12} \Dk D_{21}
	\end{array} 
	\right],
\end{align}
and the complex transfer matrices for the various frequencies $\omega$ can be obtained with 
\begin{equation}
	\Tzowo(j\omega) = \bar{C} (j\omega I - \bar{A})^{-1} \bar{B} + \bar{D}.
\end{equation} 

The robust stability criteria is $\Vert \Tzowo(s) \Vert_\infty < 1$. This can be verified by computing the maximum singular value $\bar{\sigma}$ of the complex transfer matrices $\Tzowo(j \omega)$ on a sufficiently dense frequency grid. Thus, $\crob(\bphi)$ becomes
\begin{align}
	\crob(\bphi) &= \max \Big\{1, \sup_\omega  \bar{\sigma} \big(\Tzowo(j\omega)\big)  \Big\}, \\
	\mathrm{where} \quad \bar{\sigma}(M) &= \big[\bar{\lambda} (M^*M)\big]^\frac{1}{2},
\end{align}
$M^*$ represents the conjugate transpose of a matrix $M$, and $\bar{\lambda}$ is the largest eigenvalue, which is positive real. Here again, the gradients are obtained through automatic differentiation.

\section{Example problem} \label{sec:example}
\begin{figure}
	\centering
	\includegraphics[width=0.30\linewidth,trim={2.4cm 19.0cm 14.5cm 1.6cm},clip]{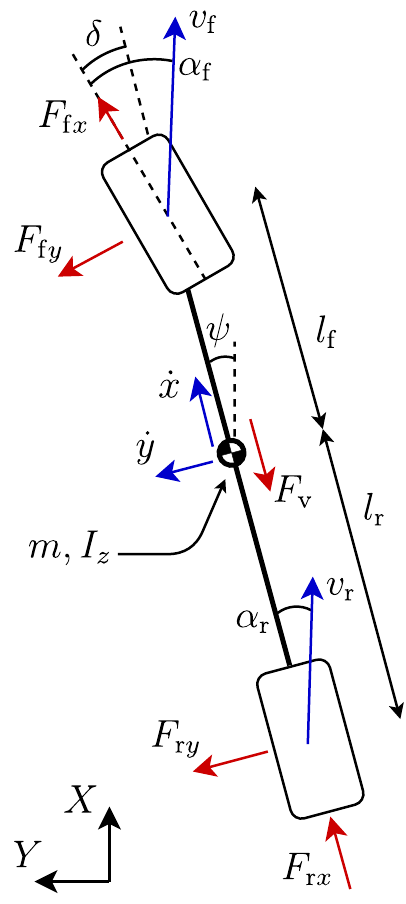}
	\caption{Free body diagram of the bicycle model.}
	\label{fig:fbd}
\end{figure}

To demonstrate the control learning method, we simulated lane-change maneuvers for autonomous vehicles~\cite{falcone_predictive_2007, liu_dynamic_2018} using the bicycle model shown in Figure~\ref{fig:fbd}. In the simulated scenarios, the vehicle is also accelerated during the lane-change maneuvers. The varying longitudinal velocity makes this a non-trivial problem and it motivates using a robust control approach~\cite{mata_robust_2019,wischnewski_tube-mpc_2022}. We assume a full-state feedback, so $\by = \bx$. The states are $\bx = [\xp, \yp, \psip, \psi, Y, \delta]^\top$ and controls are $\bu = [\Frx,\Ffx,\dr]^\top$. The system dynamics are
\begin{align}
	&f(\bx,\bu) = \nonumber \\
	&\begin{bmatrix}
		\big(\Frx +\Ffx \cos(\delta) - \Ffy \sin(\delta) - \frac{1}{2} \rho \xp^2 \Af \Cd \big)/m+ \yp \psip \ \\[3pt]
		\big(\Ffx \sin(\delta) + \Fry + \Ffy \cos(\delta)\big)/m - \xp \psip \\[3pt]
		\big(\Ffx \lf \sin(\delta) - \Fry \lr + \Ffy \lf \cos(\delta)\big)/I_z \\[3pt]
		\psip \\[3pt]
		\xp \sin(\psi) + \yp \cos(\psi) \\[3pt]
		\lambdas (\dr-\delta)
	\end{bmatrix}. \label{eq:dyn}
\end{align}
The steering behavior is modeled as a first order linear system~\cite{you_vehicle_1998}. Assuming the tire slip angles $\alphaf$ and $\alphar$ remain small, a linear tire model is used
\begin{align}
	\Ffy &= \Cf \alphaf, \quad \alphaf = \delta -\big(\psip \lf + \yp\big)/\xp,\\
	\Fry &= \Cr \alphar, \quad \alphar = \big(\psip \lr - \yp\big)/\xp,
\end{align}
where $\Cf$ and $\Cr$ are the front and rear tire stiffness coefficients. 

\begin{table}
	\centering
	\caption{Vehicle parameters}\label{tbl:params}
	\begin{tabular}{c c c c}
		Parameter & Unit                & Nominal & Uncertainty \\ \hline
		$m$    & kg                  & 2000    & $\pm 200$   \\
		$I_z$   & $\mathrm{kg\, m^2}$ & 3200    & $\pm 200$   \\
		$\Cf$ & $\mathrm{kN/rad}$   & 50      & $\pm 10$    \\
		$\Cr$ & $\mathrm{kN/rad}$   & 50      & $\pm 10$    \\
		$\lf$   & m                   & 1.1     & -           \\
		$\lr$   & m                   & 1.7     & -           \\
		$\Cd$   & -                   & 0.24    & -           \\
		$\Af$   & $\mathrm{m^2}$      & 2.4     & -           \\
		$\rho$   & $\mathrm{kg/m^3}$   & 1.225   & -           \\
		$\lambdas$   & $\mathrm{s^{-1}}$   & 8    & $\pm 2$  \\ \hline
		$\xp$   & $\mathrm{m/s}$      & 25      & +5          \\
		$\yp$   & $\mathrm{m/s}$      & 0       & $\pm 1.5$   \\
		$\psip$  & $\mathrm{rad/s}$    & 0       & $\pm 0.15$  \\
		$\psi$   & rad                 & 0       & $\pm 0.15$  \\
		$\delta$  & rad                 & 0       & $\pm 0.06$  \\
		$\Ffx$   & kN                   & 0       & -0.1/+2  \\ \hline
	\end{tabular}
\end{table}

\subsection{Linearization \& uncertainty}
The bicycle model is linearized to obtain $G(s)$. This is done by taking the Jacobian of $f(\bx,\bu)$ with respect to both states and input vectors, and evaluating the expression for the initial states $\bx_{[0]} = [\xp_0,0,0,0,0,0]^\top$ and inputs $\bu_{[0]} = [0,0,0]^\top$, thereby obtaining the $\Ag$ and $\Bg$ matrices for the state space representation of $G(s)$. Assuming full-state feedback, $\Cg = I$. \\

Input multiplicative uncertainty is used to cover the set of all possible system dynamics, where $\mathcal{G} := \big\{G(s)\big(I+W_1(s) \Delta(s) W_2(s)\big) \big\}$ and $\Vert \Delta(s) \Vert_\infty<1$. To obtain the uncertainty covers $W_1(s)$ and $W_2(s)$, multiple possible systems are sampled based on the assumed ranges for the parameters and states presented in Table~\ref{tbl:params}. Then the \textsc{Matlab} command \texttt{ucover} is used to obtain $W_1(s)$ and $W_2(s)$ such that all the sampled systems are covered by $\mathcal{G}$. 

\subsection{Trajectory \& nominal command}
The trajectories $\bxd$ consist of a concurrent lane change and vehicle acceleration. Only the states of interest $\xp$ and $Y$ are prescribed and the rest of the entries in $\bxd$ are zeros. Four scenarios are used in this work; they are parameterized with the values presented in Table~\ref{tbl:scenarios}. The vehicle is accelerated from $\xp_0$ = 25\,m/s and reaches $\xpf$ at time $\tfone$. The acceleration begins at 0\,m/s\textsuperscript{2}, then is gradually increased up to a constant value $\amax$, which is held for a moment until it is brought back to 0\,m/s\textsuperscript{2} at time $\tfone$. The lateral position begins at $Y_0$~=~0\,m and ends at $\Yf$ at time $\tftwo$. The trajectory of $Y$ follows a 3-4-5 polynomial where the first and second derivatives are null at both ends of the trajectory. \\

\begin{table}
	\centering
	\caption{Scenario description}\label{tbl:scenarios}
		\begin{tabular}{c|c c c c}
			                           & \multicolumn{4}{c}{Scenarios} \\
			        Parameter          &  1  & 2    & 3   & 4          \\ \hline
			 $\xpf \ [\mathrm{m/s}]$   & 30.0  & 26.5 & 28.0  & 30.0         \\
			 $\tfone \ [\mathrm{s}]$   & 3.0 & 3.5  & 3.5 & 5.0        \\
			$\amax \ [\mathrm{m/s^2}]$ & 2.0 & 0.5  & 1.0 & 1.8        \\
			   $\Yf \ [\mathrm{m}]$    & 3.7 & 3.7  & 3.2 & 3.8        \\
			 $\tftwo \ [\mathrm{s}]$   & 4.0 & 4.5  & 4.5 & 4.5        \\ \hline
		\end{tabular}
\end{table}

The nominal command $\bbu$ is computed in two independent steps. First, the front and rear tractive forces are computed from the reference velocity profile in $\bxd$, where
\begin{align}
	F(t) &= m\ddot{x}(t) + \tfrac{1}{2} \rho \xp(t)^2 \Af \Cd,\\
	\Ffx(t) &= \tfrac{1}{3} F(t), \quad \Frx(t) = \tfrac{2}{3} F(t). 
\end{align}
Then, the steering input is computed. A reduced dynamical system with $\Ar$ and $\Br$ is obtained by only considering the lateral dynamics in $G(s)$, i.e., only considering the states $\bx = [\yp, \psip, \psi, Y, \delta]^\top$ and controls $\bu = [\dr]$. Therefore, $\Ar$ consists of the last five columns and five rows of $\Ag$, and $\Br$ consists of the last five elements in the last column of $\Bg$. The resulting system is still controllable. The nominal steering input is then computed with  
\begin{equation}
	\dr(t) = \Br^\top \mathrm{e}^{\Ar^\top(\tftwo-t)} \Wc^{-1}(\tftwo)[0,0,0,\Yf,0]^\top,
\end{equation}
where $\Wc(t)$ is the Controllability Gramian.

\subsection{Controller structure and initialization}
The feedback controller is structured to decouple the control of the longitudinal and lateral dynamics. The front $\Ffx$ and rear $\Frx$ tractive forces are only used for the feedback control of the vehicle velocity $\xp$, while the steering input $\dr$ is only used for controlling the lateral dynamics. As will become apparent in the result section, a simple integrator is sufficient for the vehicle speed control. Moreover, the tracking error is so negligible that no further controller tuning is required for this integrator. \\

The control of the lateral dynamics is more difficult, however. An initial controller is obtained with $\Hinf$ loop shaping. For that, a nominal linear model $G(s)$ is built from the same $\Ar$ and $\Br$ matrices introduced in the previous section. Then, a controller that stabilizes $G(s)$ and minimizes the tracking error is obtained using the \textsc{Matlab} command \texttt{hinfsyn}. This $\Hinf$ controller is assembled with the simple integrator obtained previously to form the complete initial feedback controller $\Ki$, which has 24 internal states, and has $\Dk = 0$. Together the $\Ak$, $\Bk$, and $\Ck$ matrices of $\Ki$ contain 792 entries.

\subsection{Controller training with nominal model}
The controller $\Kt$ is tuned with simulated rollouts, see line~\ref{algl:while} in Algorithm~\ref{alg:RobStab}. The system model $\fn(\bx,\bu)$ is that of Equation~\eqref{eq:dyn} with the nominal parameters shown in Table~\ref{tbl:params}. The simulations have a time step of $\Ts$ = 0.02\,s. The cost function used is
\begin{equation}
	c(\bx_{[t]}) = (\bx_{[t]} - {\bxd}_{[t]})^\top Q (\bx_{[t]} - {\bxd}_{[t]}),
\end{equation}
where $Q$ is a diagonal matrix weighting the relative importance of the state errors. \\

Only the first three scenarios of Table~\ref{tbl:scenarios} are used during training. Scenario 4 is reserved for testing that the controller also does well on a scenario never seen during training.

\subsection{Controller training with learned model}\label{sec:learned_model}
The controller $\Ka$ is also tuned from simulated policy rollouts, but with a learned model of the system dynamics this time, see line~\ref{algl:while2} in Algorithm~\ref{alg:RobStab}. In this example problem, the actual (unknown) system has an altered tire stiffness coefficient of $\Cf$ = $\Cr$ = 40\,kN/rad. We chose to learn the dynamics using a model that has the form 
\begin{equation}
	\fl = \fn + [g_1(\bz),g_2(\bz),g_3(\bz),0,0,0]^\top
\end{equation}
where $g_d(\bz)$ are functions that represent the unknown dynamics and $\bz = [\alphaf,\alphar]^\top$ is the feature vector used in the prediction models. To learn $\fl$, first a lane change is simulated for Scenario 1 with $\Kt$ (line~\ref{algl:imp} in Algorithm~\ref{alg:RobStab}), and a dataset $\mathcal{D} = \{\bx_{[i]},\bu_{[i]}\}$ composed of $\np$ + 1 data points is collected. Then Gaussian processes are used to learn the difference between the nominal system dynamics $\fn$ and the actual dynamics. In other words, the targets to be learned are issued from the prediction error
\begin{equation}
	{\bep}_{[i]} = \bx_{[i+1]} - \fn(\bx_{[i]},\bu_{[i]}).
\end{equation}
More specifically, each of the three dimensions to be learned has its own target vector $\td \in \mathbb{R}^{\np}$, which are composed of the $d$-th component of the error vectors, such that
\begin{equation}
	\td = \big[[{\bep}_{[1]}]_d,\ldots,[{\bep}_{[\np]}]_d\big]^\top.
\end{equation}
All three dimensions share the same set of feature vectors $\{ \bz_{[i]} \}$. The Gaussian processes used in this work has a zero mean function $m(\bz) := 0$ and a square exponential kernel function
\begin{equation}
	k(\bz_i,\bz_j) := \sigmaf^2 \exp(-\tfrac{1}{2}(\bz_i - \bz_j)^\top \Lambda^{-1}(\bz_i - \bz_j)),
\end{equation}
where $\sigmaf^2$ is the signal variance, and $\Lambda = \mathrm{diag}([l_1^2,l_2^2])$ is a diagonal matrix composed of the characteristics length-scales of the two dimensions in $\bz$. These are the hyper-parameters of the Gaussian process, and every function $g_d$ has its own set of hyper-parameters. The predictive functions $g_d(\bz)$ are simply the mean of the Gaussian processes evaluated at the test inputs~$\bz_*$
\begin{equation}
	g_d(\bz_*) = \mu_d(\bz_*) = \Kzsz(\Kzz + \sigma_\epsilon^2 I)^{-1} \td, \label{eq:mu}\\
\end{equation}
where $\sigma_\epsilon^2$ is the noise variance, $\Kzz = [k(\bz_i,\bz_j)]_{ij}$, and $\Kzsz = [k(\bz_*,\bz_i)]_i$. The hyper-parameters, symbolically regrouped in the vector $\btheta$, can be tuned by maximizing the logarithm of the marginal likelihood of the observed data points in $\mathcal{D}$, as suggested in~\cite{rasmussen_gaussian_2006}, with
\begin{equation}
	\log p(\td \vert \mathcal{D},\btheta) = -\frac{1}{2} \td^\top K^{-1} \td - \frac{1}{2} \log \vert K \vert - \frac{n}{2} \log 2 \pi,
\end{equation}
where $K = \Kzz + \sigma_\epsilon^2 I$, and $\vert K \vert$ is the determinant of $K$. The likelihood is maximized with a gradient-based optimization algorithm, where $\balpha = K^{-1} \td$,
\begin{equation}
	\dpd{}{\btheta_j} \log p(\td \vert \mathcal{D},\btheta) = \frac{1}{2} \tr \bigg( \big( \balpha \balpha^\top - K^{-1}\big) \dpd{K}{\btheta_j} \bigg).
\end{equation}

\section{Results \& discussion} \label{sec:results}
\begin{figure}
	\centering
	\subfloat[Vehicle speed, Scenario 1.]{\includegraphics[width=0.45\linewidth,trim={0.0cm 0cm 0.5cm 0cm},clip]{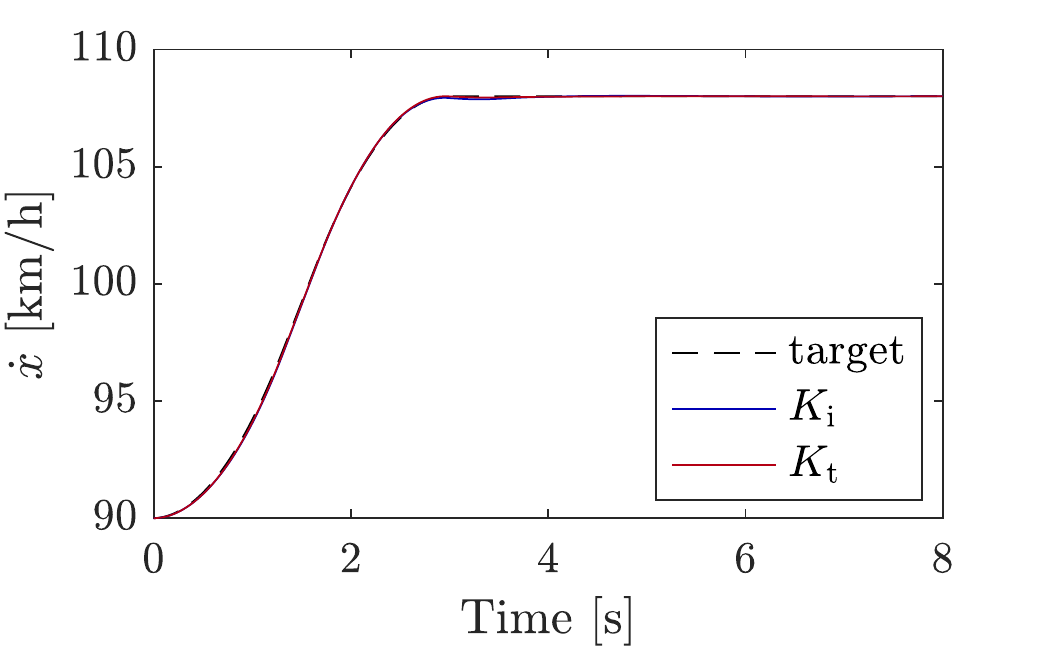}}	\quad
	\subfloat[Lateral position, Scenario 1. \label{fig:Yplot1_traj1}]{\includegraphics[width=0.45\linewidth,trim={0.0cm 0cm 0.5cm 0cm},clip]{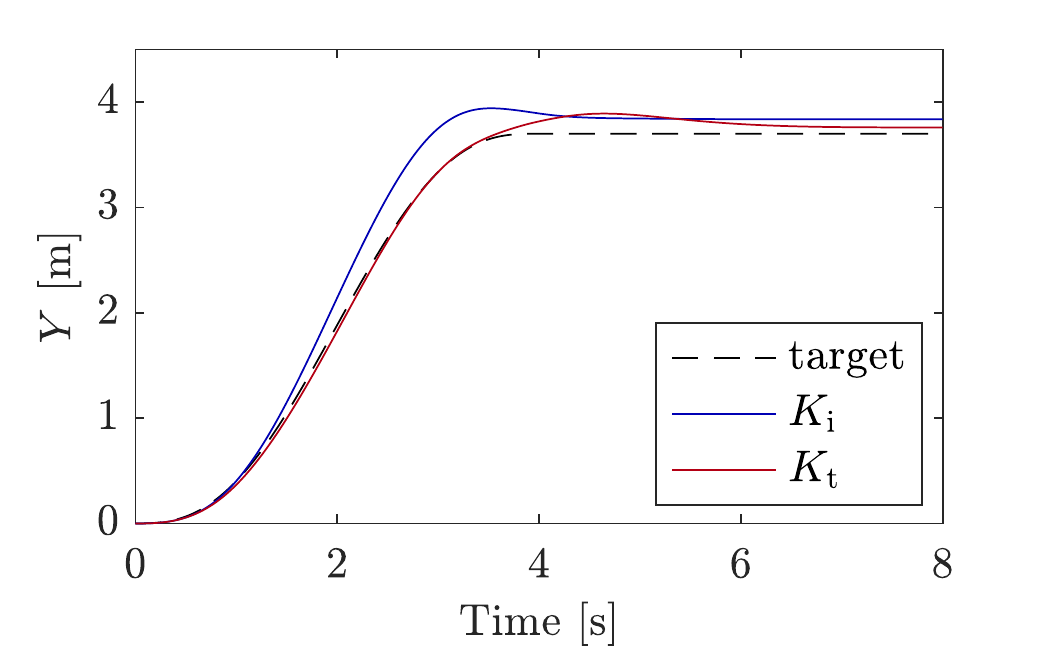}}
		
	\caption{Simulated policy rollouts on the nominal model ($\Cf$ = $\Cr$ = 50\,kN/rad). The tuned controller $\Kt$ improves the lateral position tracking compared to initial controller $\Ki$.}
	\label{fig:results_1}
\end{figure}

Figure~\ref{fig:results_1} shows that the tracking error is negligible for the vehicle speed $\xp$, but not for the lateral position $Y$. Therefore, the result section only discusses the tracking of the lateral position. \\

\begin{figure}
	\centering
	\subfloat[Lateral position, Scenario 1. \label{fig:eYplot1_traj1}]{\includegraphics[width=0.45\linewidth,trim={0.0cm 0cm 0.5cm 0cm},clip]{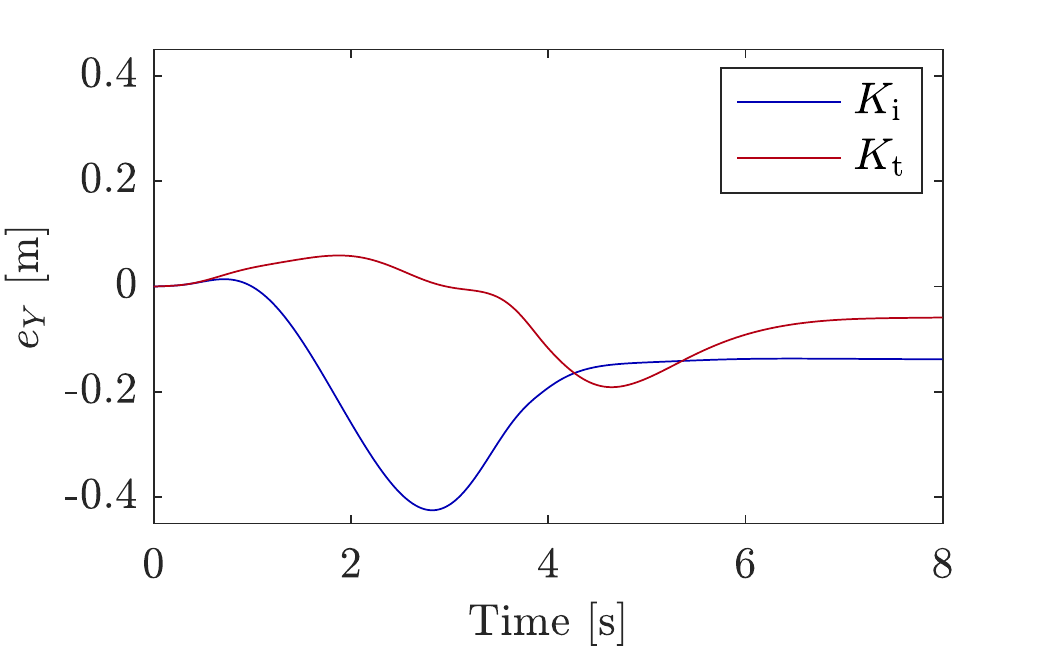}}	 \quad
	\subfloat[Lateral position, Scenario 4. \label{fig:eYplot1_traj4}]{\includegraphics[width=0.45\linewidth,trim={0.0cm 0cm 0.5cm 0cm},clip]{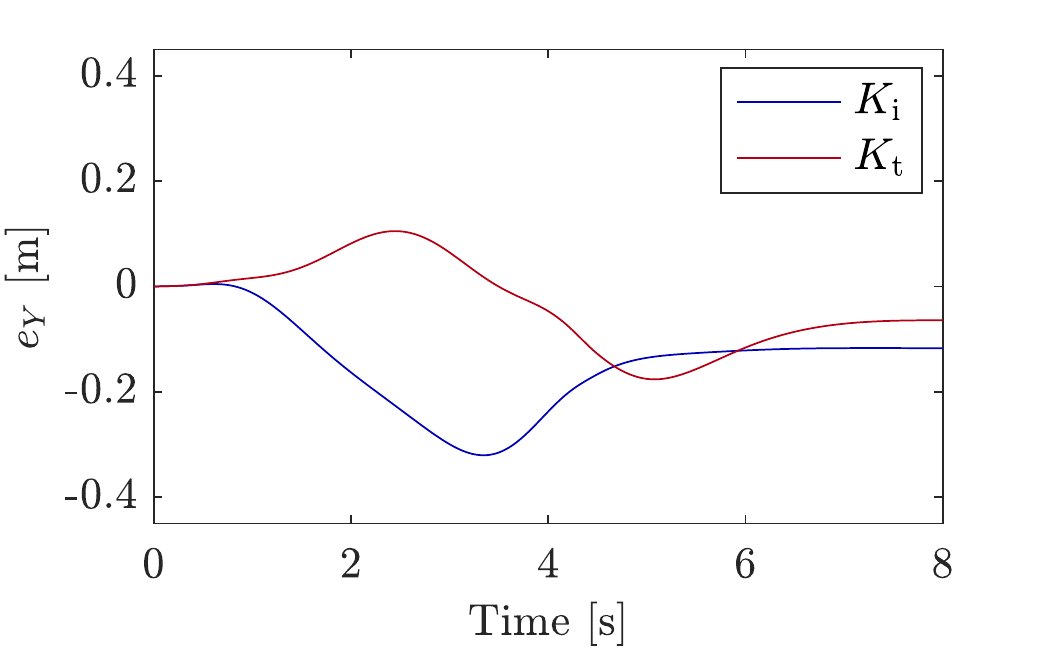}}

	\caption{Simulated policy rollouts on the nominal model ($\Cf$ = $\Cr$ = 50\,kN/rad). The tuned controller $\Kt$ improves the tracking performance for both a scenario used during training (Scenario~1) and a scenario not used during training (Scenario~4).}
	\label{fig:results_2}
\end{figure}

\begin{figure}
	\centering
	\includegraphics[width=0.50\linewidth,trim={0cm 0cm 0.5cm 0cm},clip]{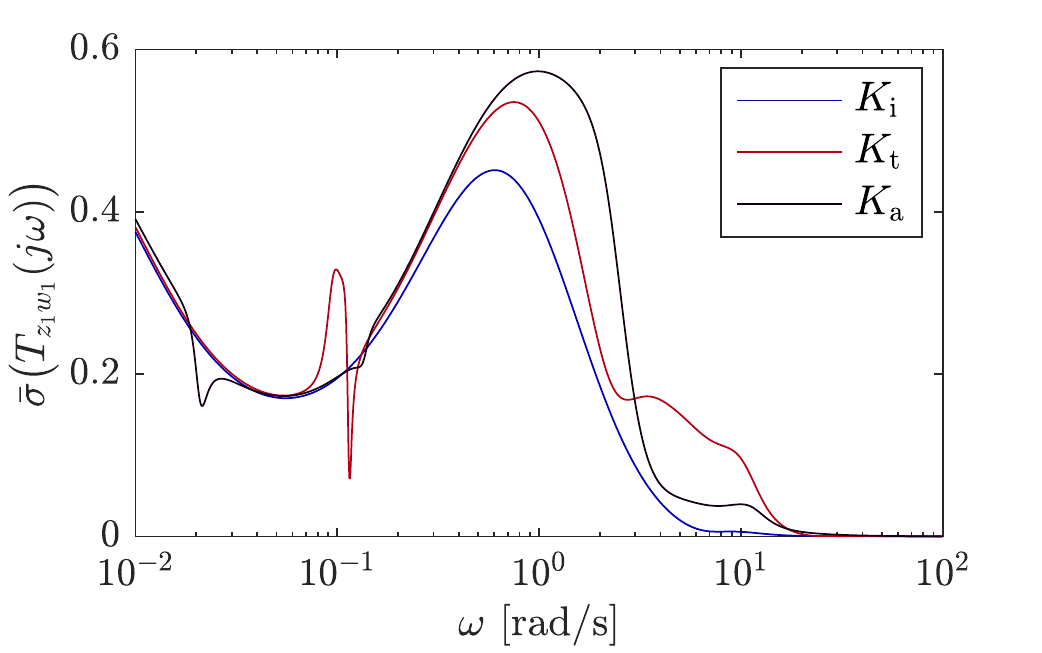}
	\caption{The maximum singular value $\bar{\sigma} \big( \Tzowo(j\omega) \big)$ for $\Ki$, $\Kt$, and $\Ka$. }
	\label{fig:muplot}
\end{figure}

Figure~\ref{fig:Yplot1_traj1} shows that the controller tuned from simulated policy rollouts ($\Kt$) improves the tracking performance when compared to the initial controller ($\Ki$) obtained from $\Hinf$ loop-shaping. Figure~\ref{fig:eYplot1_traj1} shows the tracking error $e_Y$ on the lateral position for this run, and Table~\ref{tbl:error} presents the $\mathcal{L}_2$-norm of this error signal. The tuning method also reduces the tracking error for a lane-change scenario that was not used during training. In effect, both Figure~\ref{fig:eYplot1_traj4} and Table~\ref{tbl:error} show that $\Kt$ does better than $\Ki$ for Scenario~4. This suggests that the tuning method does not overfit on the training scenarios, and can generalize to others. From experience, this was not the case when a single scenario was used for training. Finally, the proposed method keeps the tuned controller $\Kt$ robustly stabilizing, which can be seen in Figure~\ref{fig:muplot}.\\

\begin{table}
	\centering
	\caption{$\mathcal{L}_2$-norm of the tracking error on $Y$ ($e_Y$) for the 0-8\,s span.}\label{tbl:error}
		\begin{tabular}{c|c c c c}
			& \multicolumn{4}{c}{Scenarios} \\
			\textbf{Nominal model}  &  1   & 2    & 3    & 4        \\ \hline
			$\Ki$           & 3.85 & 2.22 & 2.92 & 3.39     \\
			$\Kt$           & 1.78 & 1.53 & 1.33 & 1.80     \\ \hline
			\textbf{Actual dynamics} &      &      &      &          \\ \hline
			$\Kt$           & 3.84 & 3.85 & 3.11 & 3.87     \\
			$\Ka$           & 2.21 & 2.55 & 1.90 & 2.27     \\ \hline
	\end{tabular}
\end{table}

\begin{figure}
	\centering
	\subfloat[Performance cost $\cp$. \label{fig:perf}]{\includegraphics[width=0.45\linewidth,trim={0.0cm 0cm 0.4cm 0cm},clip]{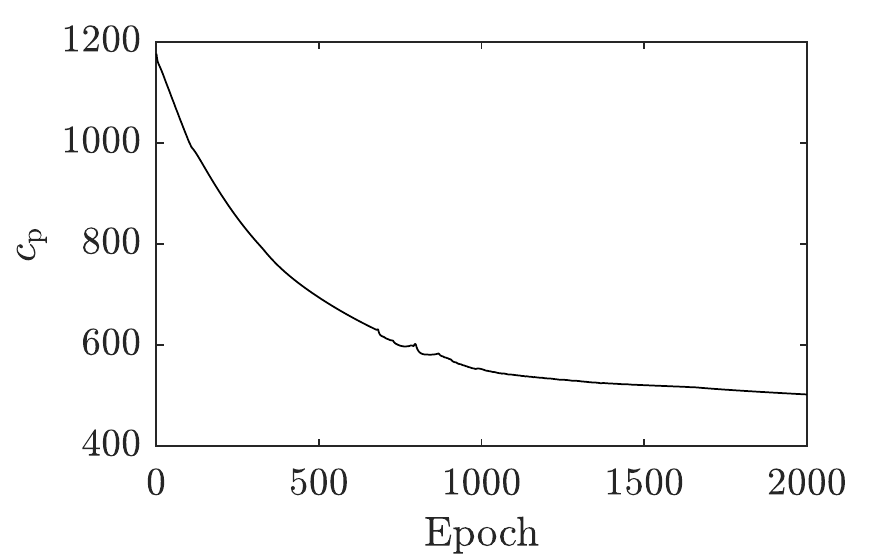}} \\
	\subfloat[Nominal stability criteria. \label{fig:stab}]{\includegraphics[width=0.45\linewidth,trim={0.0cm 0cm 0.4cm 0cm},clip]{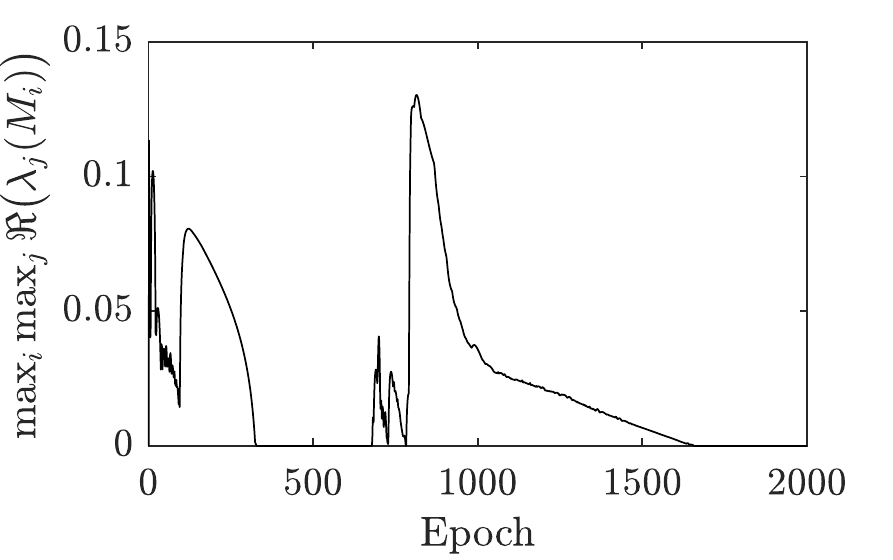}} \quad
	\subfloat[Robust stability criteria. \label{fig:rob}]{\includegraphics[width=0.45\linewidth,trim={0.0cm 0cm 0.4cm 0cm},clip]{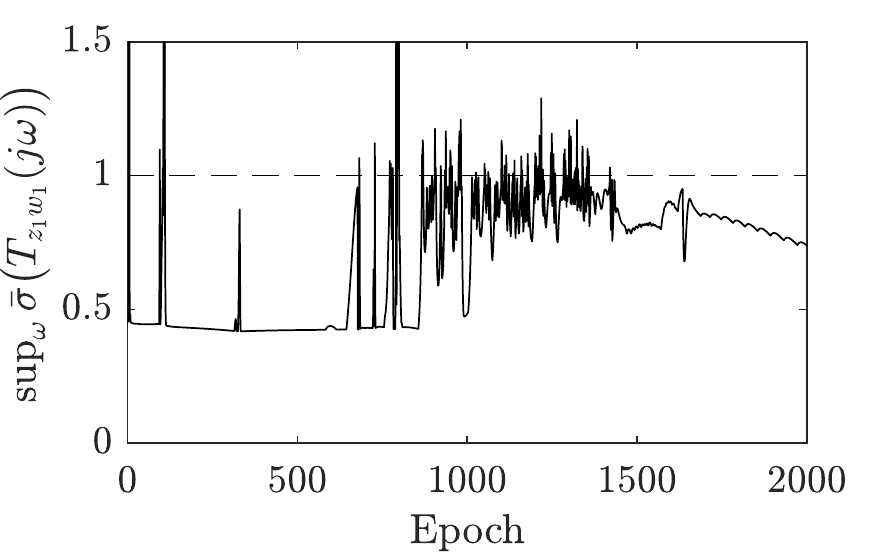}}

	\caption{Evolution of the performance cost, nominal stability criteria, and robust stability criteria during the first 2000 training epochs for $\Kt$.}
	\label{fig:epochs}
\end{figure}

The tuning of $\Kt$ requires 4000 training epochs. The first 2000 epochs are done with simulated rollouts of 8\,s duration (simulated time), and the last 2000 epochs, with 40\,s rollouts. The longer rollouts are used to ensure that the low-frequency dynamics are also considered during the controller tuning. The total training time for the first 2000 epochs is 1083\,s, and for the last 2000 epochs, 5245\,s. Figure~\ref{fig:epochs} shows the evolution of the performance cost $\cp$ in (a), the maximum real part of all eigenvalues in the matrices $M_1$ to $M_4$ in (b), and the maximum singular value of $\Tzowo$ in (c). This indicates that the penalties $\xis \cs$ and $\xir \crob$ were sufficient to enforce both the nominal and robust stability constraints. Figure~\ref{fig:epochs} shows the training behavior when the weights are $\xis$ = 10,000 and $\xir$~=~1000. For smaller values of $\xis$ and $\xir$ however, the penalties were not strong enough to enforce the stability constraints. The gradient descend algorithm used during training is Adam~\cite{kingma_adam_2017}, which helps stabilize the gradients compared to a simple stochastic gradient descent algorithm. \\

\begin{figure}
	\centering
	\subfloat[Lateral position, Scenario 1. \label{fig:eYplot2_traj1}]{\includegraphics[width=0.45\linewidth,trim={0.0cm 0cm 0.5cm 0cm},clip]{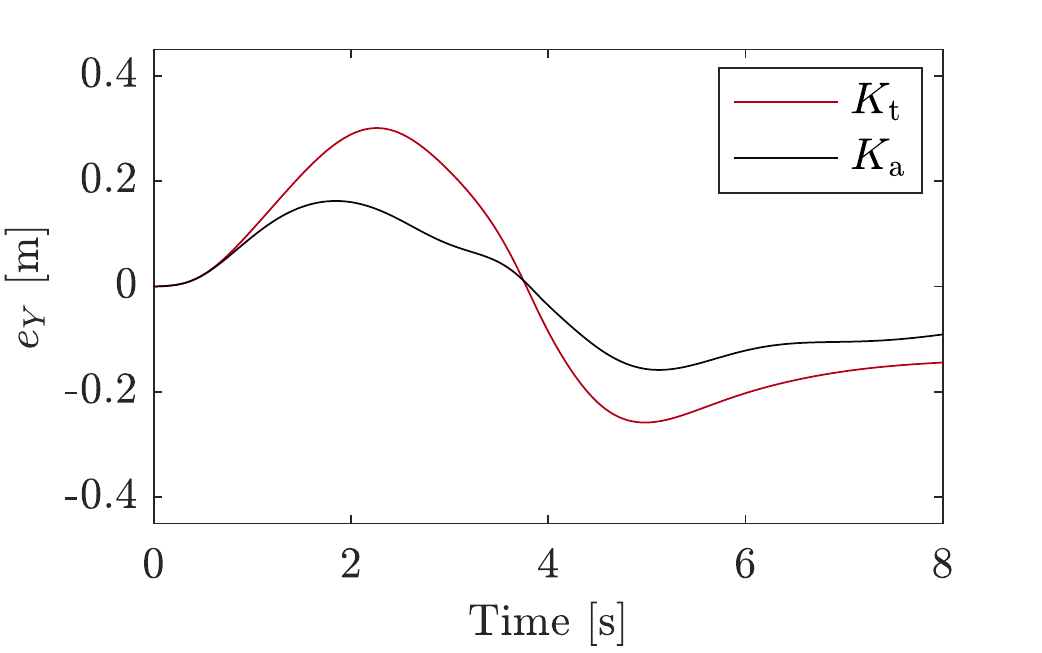}} \quad
	\subfloat[Lateral position, Scenario 4. \label{fig:eYplot2_traj4}]{\includegraphics[width=0.45\linewidth,trim={0.0cm 0cm 0.5cm 0cm},clip]{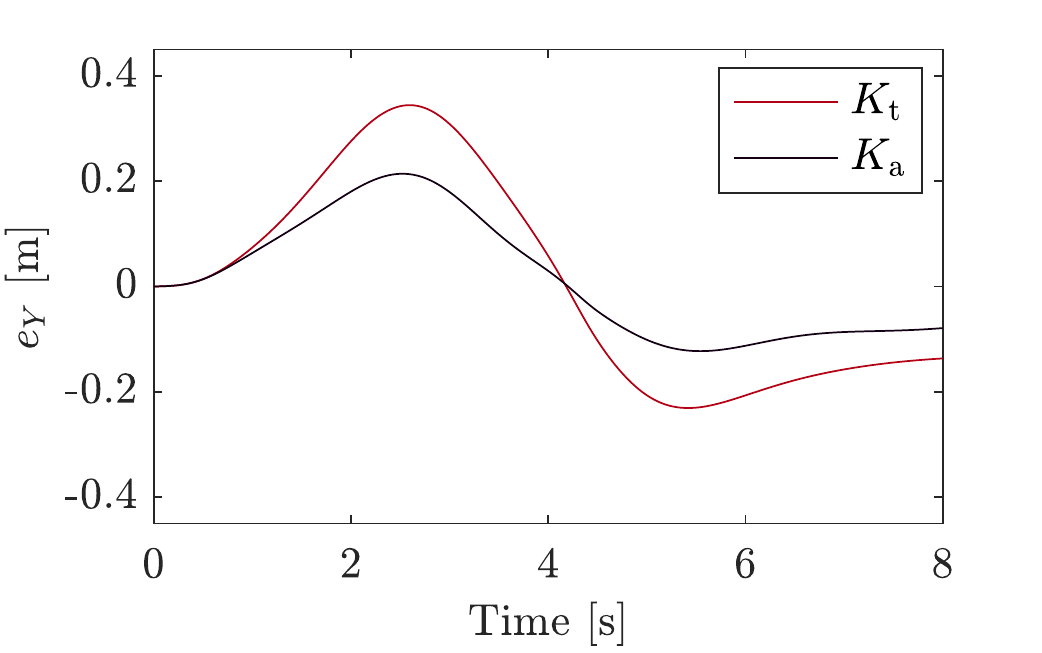}}

	\caption{Simulated policy rollouts on the actual system dynamics ($\Cf$ = $\Cr$ = 40\,kN/rad). The controller $\Ka$ has a better tracking performance than $\Kt$. This is because $\Ka$ was tuned on a learned model $\fl$ while $\Kt$ was only tuned on a nominal model $\fn$.}
	\label{fig:results_3}
\end{figure}

Figure~\ref{fig:results_3} and Table~\ref{tbl:error} show that the tracking error of $\Kt$ increases for policy rollouts on the actual dynamics, i.e. when $\Cf$ = $\Cr$ = 40\,kN/rad. However, the controller remains stable, as the actual dynamics are still within the assumed uncertainty bounds. This allows to bring back the performance to a similar level by re-tuning the controller with a learned model. As a result, the adjusted controller $\Ka$ reduces the tracking error, even for the unseen Scenario~4. Training with the learned model is longer, but still reasonable as the same 2000 epochs of 8\,s rollouts takes 2071\,s of training time, which is approximately twice the time for the nominal model. 

\subsection{Augmented cost}
It is possible to augment the total cost $J(\bphi)$ of Equation~\eqref{eq:cost} with ${\ct(\bphi) = \Vert \Tztwt(s) \Vert_\infty}$ to help shape the closed-loop behavior with frequency-based specifications. The gradients $\nabla_{\bphi}\ct(\bphi)$ can be computed with the method of Section~\ref{sec:rob_grad} used to obtain $\nabla_{\bphi}\crob(\bphi)$, but with $\Tztwt(j\omega)$ instead. This additional cost $\ct$ could be used in problems where scenario-based tuning alone is inadequate. One such example could be a controller that must account for a large frequency range. In effect, if $\Ts$ is small enough to allow the simulation of high-frequency dynamics, it also means that the policy rollouts must have a long time span to also include the effects of low-frequency dynamics. Therefore, the cost $\ct$ could be used to penalize the low-frequency errors, while the scenario-based performance cost $\cp$ tunes the high-frequency behavior. In the lane-change problem however, this was found unnecessary. As previously discussed, it was sufficient to simulate 40\,s rollouts. As a final remark, the nominal model $G(s)$ is not learned in the proposed method, so the additional cost $\ct$ could be less beneficial if the actual system deviates significantly from $G(s)$. 

\section{Conclusion}
This article presents a method to improve the performance of a LTI controller from a learned nonlinear model, while preserving robust stability using a robust control framework. The controller parameters are iterated with gradients obtained from simulated rollouts with the learned model. The robust stability criteria are enforced through the gradients as well, penalizing only unsafe controllers. The method yields a tuned controller that is already robustly stabilizing when implemented on the actual system for the first time. It also remains stable when sudden changes in the system dynamics occur, given that they are within assumed uncertainty bounds. This was demonstrated by simulating a bicycle model performing lane-change maneuvers. The approach successfully learns a robustly stabilizing controller, and thereby takes advantage of modern machine learning for synthesizing LTI controllers. 

\bibliographystyle{IEEEtranMdoi}
\bibliography{main}

\begin{IEEEbiography}[{\includegraphics[width=1in,height=1.25in,clip,keepaspectratio]{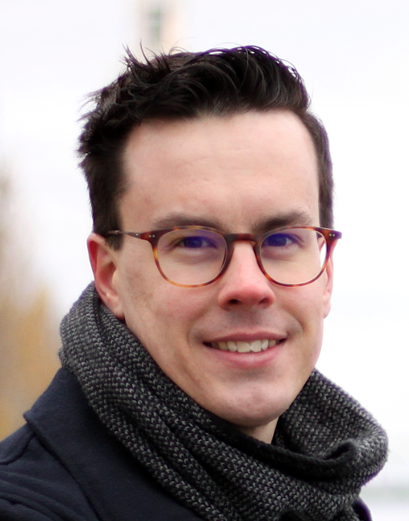}}]{Marc-Antoine Beaudoin}
	is a doctoral student at the Intelligent Automation Laboratory of the Centre for Intelligent Machines at McGill University. He obtained a Bachelor of Engineering degree from Université de Sherbrooke in 2015, and a Master of Applied Science degree from the University of Toronto in 2018, both in Mechanical Engineering. He is currently pursuing the Ph.D. degree in Electrical and Computer Engineering at McGill University. Prior to his doctoral studies, Mr. Beaudoin was a practicing engineer in the electric vehicle industry. His research areas include the machine learning control of dynamical systems, with a focus on electric and autonomous vehicles.
\end{IEEEbiography}
\begin{IEEEbiography}[{\includegraphics[width=1in,height=1.25in,clip,keepaspectratio]{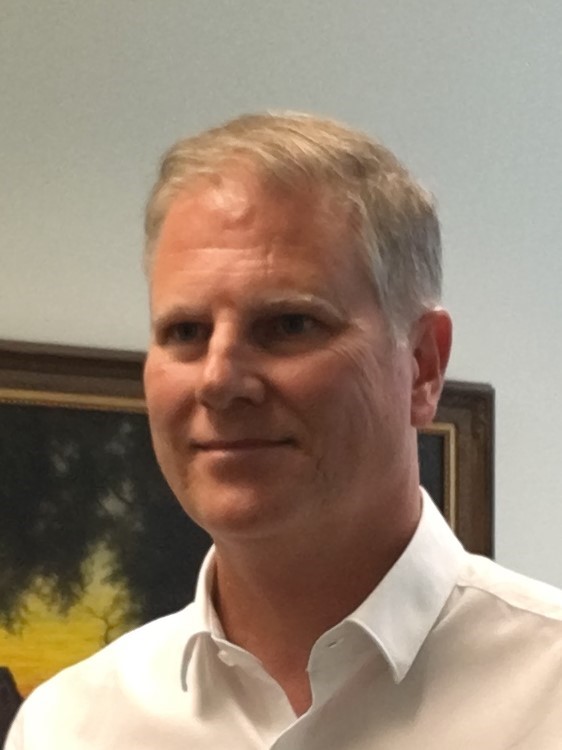}}]{Benoit Boulet}
	(Senior Member, IEEE) is Professor in the Department of Electrical and Computer Engineering at McGill University which he joined in 1998, and Director of the McGill Engine, a Technological Innovation and Entrepreneurship Centre. He is Associate Vice-Principal of McGill Innovation and Partnerships and was Associate Dean (Research \& Innovation) of the Faculty of Engineering from 2014 to 2020. Professor Boulet obtained a Bachelor's degree in applied sciences from Université Laval in 1990, a Master of Engineering degree from McGill University in 1992, and a Ph.D. degree from the University of Toronto in 1996, all in electrical engineering. He is a former Director and current member of the McGill Centre for Intelligent Machines where he heads the Intelligent Automation Laboratory. His research areas include the design and data-driven control of electric vehicles and renewable energy systems, machine learning applied to biomedical systems, and robust industrial control.
\end{IEEEbiography}

\end{document}